\documentclass{Interspeech}

\interspeechcameraready

\title{Investigation of Zero-shot Text-to-Speech Models for Enhancing Short-Utterance Speaker Verification}

\author[affiliation={1}]{Yiyang}{Zhao}
\author[affiliation={2}]{Shuai}{Wang}
\author[affiliation={3}]{Guangzhi}{Sun}
\author[affiliation={1}]{Zehua}{Chen}
\author[affiliation={1}]{Chao}{Zhang}
\author[affiliation={1}]{\\Mingxing}{Xu}
\author[affiliation={1*}]{Thomas Fang}{Zheng}

\address{
  $^1$Tsinghua University, China,
  $^2$Shenzhen Research Institute of Big Data, China \\
  $^3$University of Cambridge, United Kingdom\thanks{\hspace{-1ex}*Corresponding author}}
\email{ zhaoyy22@mails.tsinghua.edu.cn, wangshuai@cuhk.edu.cn, gs534@cam.ac.uk, \{zhc23thuml,cz277,xumx,fzheng\}@tsinghua.edu.cn}


\usepackage{xcolor}

\usepackage{multirow}
\usepackage{cite}
\usepackage{adjustbox}
\usepackage{hyperref}

\begin{document}

\nolinenumbers

\maketitle

\begin{abstract}
Short-utterance speaker verification presents significant challenges due to the limited information in brief speech segments, which can undermine accuracy and reliability. Recently, zero-shot text-to-speech (ZS-TTS) systems have made considerable progress in preserving speaker identity. In this study, we explore, for the first time, the use of ZS-TTS systems for test-time data augmentation for speaker verification. We evaluate three state-of-the-art pre-trained ZS-TTS systems, NatureSpeech 3, CosyVoice, and MaskGCT, on the VoxCeleb 1 dataset. Our experimental results show that combining real and synthetic speech samples leads to 10\%–16\% relative equal error rate (EER) reductions across all durations, with particularly notable improvements for short utterances, all without retraining any existing systems. However, our analysis reveals that longer synthetic speech does not yield the same benefits as longer real speech in reducing EERs. These findings highlight the potential and challenges of using ZS-TTS for test-time speaker verification, offering insights for future research.

\end{abstract}

\section{Introduction}

Speaker verification (SV) aims to confirm an individual’s identity based on their voice characteristics. In recent years, deep neural network (DNN)-based models \cite{snyder2018x, desplanques2020ecapa,zeinali2019but, zhang2022mfa, gu2023memory, zhao2024whisper, wang2023cam++, chen2024eres2netv2} have driven remarkable advancements in SV systems.
Despite these achievements, SV systems still encounter challenges in short-utterance scenarios~\cite{wang2024overview}. Unlike longer speech samples, which may last several seconds or even minutes, short-utterance speech contains limited information and is highly susceptible to interference from factors such as content and channel effects. To address these challenges, researchers have proposed various solutions, including designing more efficient aggregation methods \cite{hajavi2019deep, tang2019deep, jung2020improving, gao2019improving}, explicitly aligning short and long speech representations \cite{kye2020meta, liu2020text,jung2019short,sang2020open}, and predicting long-term embeddings from short-term ones \cite{guo2017cnn, zhang2018vector}.

Some related studies have explored the use of text-to-speech (TTS) systems to enhance SV systems. These methods can be categorized into two main approaches: (1) expanding the diversity of training data by synthesizing new speakers\cite{wang2020data}; and (2) generating synthetic speech with specific text for text-dependent speaker verification tasks\cite{qin2020exploring,du2021synaug}.

Notably, the above approaches focus on synthesizing additional training data to improve the performance of the speaker encoder. However, the application of TTS systems for test-time augmentation remains unexplored. This is primarily due to the limited data available from a single speaker during testing, which is unseen in the training set. Such scenarios require TTS systems to exhibit strong zero-shot cloning capabilities. Traditional TTS models, however, often struggle to preserve speaker identity reliably in zero-shot settings.

Recent advancements in zero-shot TTS have brought about significant progress, with the emergence of systems such as VALL-E~\cite{wang2023neural}, CosyVoice~\cite{du2024cosyvoice}, and F5-TTS\cite{chen2024f5}, which have garnered significant attention for their high-quality synthesis and robust zero-shot cloning capabilities. Many of these systems claim to achieve highly accurate voice cloning with just a few seconds of prompt audio, preserving not only timbre and prosody but even certain acoustic environments present in the prompt. 
In this context, we are curious whether these advanced ZS-TTS systems can effectively assist in short-utterance speaker verification tasks. 

In this paper, we present the first systematic investigation of ZS-TTS systems' capabilities in short-utterance speaker verification scenarios, aiming to answer four fundamental questions:

\begin{itemize}[left=0pt, label=\textbullet ]  
    \item \textbf{Do different TTS architectures exhibit similar performance patterns?} No, performance varies: CosyVoice and NaturalSpeech 3 excel with shorter prompts, while MaskGCT performs better with longer prompts.
    \item \textbf{How critical is the selection content to generate for speech augmentation?} We found that using both original transcripts and consistent text content prove effective. However, the phoneme coverage in the synthesized speech does influence verification accuracy.
    \item \textbf{How does TTS augmentation efficacy vary with speech duration?} With fusion, TTS provides more significant relative improvements for very short speech, though overall performance remains poor. For longer speech, the benefit is less pronounced. Additionally, TTS stability decreases significantly for when only short prompts are available.
    \item \textbf{Can zero-shot TTS-generated speech enhance short-duration SV systems?}  Yes, but not as replacement of the original speech. While direct synthesis of extended speech using prompts from test utterances shows minimal impact, fusion of embeddings from TTS-generated and original speech demonstrates measurable performance gains in short-duration speaker verification.
\end{itemize}

\begin{figure*}[htbp]
  \centering
  \includegraphics[width=0.9\textwidth]{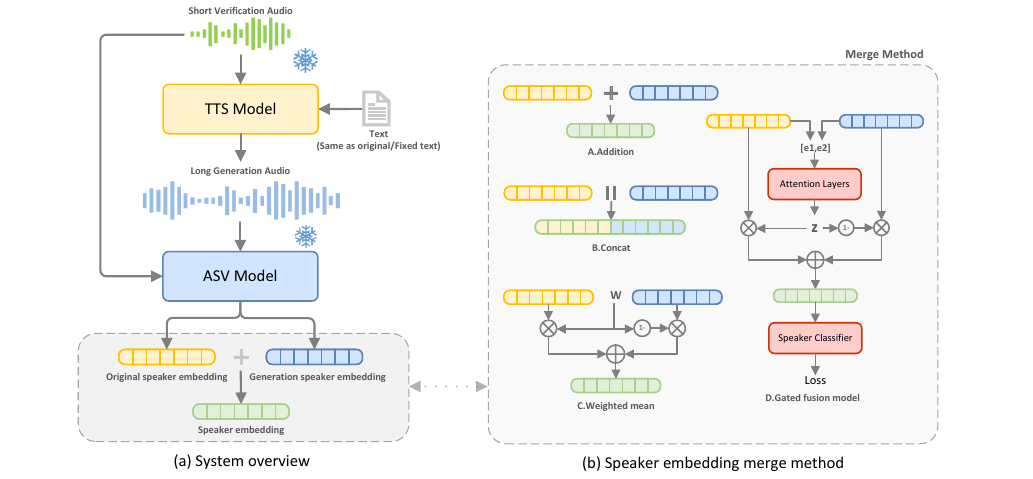}
  \caption{TTS-Enhanced Short-Utterance Speaker Recognition Pipeline.}
  \label{fig:figure1}
\end{figure*}

In summary, due to the limitations of TTS models in maintaining speaker consistency with short prompts, the improvements remain constrained. We hope our research provides valuable insights and practical guidance for leveraging TTS to enhance short-duration speaker recognition tasks.\footnote{It should be noted that all conclusions are based on experiments conducted on VoxCeleb1, which itself is not entirely clean and may increase the difficulty of TTS cloning.}




\section{Method}

In this section, we describe our overall design, which leverages TTS-generated speech to provide additional information for short-utterance speaker recognition, without modifying the original speaker model in any way. Figure~\ref{fig:figure1} illustrates the entire workflow along with implementation details.

\subsection{Speaker Embedding Extraction}

We begin by using the original bona fide short utterance \( x_b \) as a prompt to generate a longer utterance \( x_g \) through a TTS model. Both the original short speech and the TTS-generated speech are then fed into the speaker recognition network (ASV) to extract their respective speaker embeddings. Formally,

\[
\begin{gathered}
x_g = \mathcal{T}(x_b, t) \\
\mathbf{e}_b = \mathcal{F}(x_b), \quad \mathbf{e}_g = \mathcal{F}(x_g)
\end{gathered}
\]

where \( \mathcal{T}(\cdot) \) denotes the TTS model, \( x_b \) is the original short speech, and \( x_g \) is the TTS-generated speech. \( \mathcal{F}(\cdot) \) represents the ASV model that extracts the speaker embeddings \( \mathbf{e}_b, \mathbf{e}_g \in \mathbb{R}^d \). The variable \( t \) denotes the text prompt provided to the TTS model for speech synthesis. In our experiments, \( t \) can be controlled in two ways:

\begin{itemize}
    \item \textbf{Fixed text}: Synthesizing the same sentence for all test utterances. To mitigate the impact of text content randomness, the results are averaged over three different fixed texts.
    \item \textbf{Original text}: Using the original (before segmentation) transcript for each utterance.
\end{itemize}

By utilizing the TTS-generated longer utterance \( x_g \), we aim to enrich the amount of speaker-relevant information available for embedding extraction, which can potentially improve speaker verification performance in short-utterance scenarios.

In the experiment validating the phoneme impact, we used GPT-4o to assist in text generation and verified it with the CMU Pronouncing Dictionary from the NLTK library. The text prompts were carefully controlled to have similar lengths while varying the number of phonemes included in the text.

Note that the TTS augmentation is applied only to the enrollment and test audios, and is not used to augment the training data. Therefore, the speaker model itself remains unchanged.

\subsection{Embedding Fusion Methods}
In our experiments, we found that directly replacing the original bonafide short speech embedding with the embedding extracted from the generated speech leads to a performance drop. Therefore, we explored how to use both simultaneously and experimented with different fusion strategies.
As shown in Figure~\ref{fig:figure1}(b), we explore various embedding fusion strategies to combine the $\mathbf{e}_b$ and the TTS-generated speaker embedding $\mathbf{e}_g$. By leveraging the complementary information in these two embeddings, we aim to improve the performance of short-utterance speaker recognition. The commonly adopted methods such as addition, concatnation and weighted mean are as follows,

\begin{equation}
\mathbf{e}_s = 
\begin{cases} 
    \mathbf{e}_b + \mathbf{e}_g, & \text{Addition} \\ 
    [\mathbf{e}_b, \mathbf{e}_g], & \text{Concatenation} \\ 
    w \cdot \mathbf{e}_b + (1 - w) \cdot \mathbf{e}_g, & \text{Weighted Mean} 
\end{cases}
\end{equation}

Furthermore, we also tried to learn a parameterized attention-based fusion: given $\mathbf{e}_b$ and $\mathbf{e}_g$, we first compute a gate vector $\mathbf{z} \in \mathbb{R}^d$ as

\begin{equation}
\mathbf{z} = \sigma\big(\text{ATT}(\mathbf{e}_b, \mathbf{e}_g)\big)
\end{equation}

where $\sigma(\cdot)$ is the sigmoid function, we then fuse $\mathbf{e}_b$ and $\mathbf{e}_g$ via

\begin{equation}
\mathbf{e}_s = \mathbf{z} \odot \mathbf{e}_b + \big(1 - \mathbf{z}\big) \odot \mathbf{e}_g,
\end{equation}

Finally, we adopt the same speaker classifier and loss function used in the original speaker model, initializing their parameters from the pre-trained speaker model.

\section{Experimental Setup}

\subsection{Dataset}

To investigate the effectiveness of the proposed approach, we conduct experiments on the VoxCeleb1 \cite{nagrani2020voxceleb} dataset, which contains over 150,000 utterances from 1,251 celebrities. This dataset spans diverse ethnicities and accents, making it a challenging benchmark for speaker recognition. 
Since our objective is to evaluate how TTS systems can aid short-utterance speaker recognition, rather than to achieve state-of-the-art performance, we opt for the smaller VoxCeleb1 dataset instead of VoxCeleb2 for fast experimental validation.

For data preprocessing, we modify the original test set to simulate short-utterance scenarios. Specifically, each test utterance is truncated at its midpoint to produce a single segment with a duration of 0.5 seconds, 1 second, 2 seconds, or the full length of the utterance. This approach minimizes the occurrence of empty segments while allowing us to systematically evaluate the impact of varying utterance lengths on system performance.

\subsection{Speaker Recognition Model}

We adopt the ECAPA-TDNN\cite{desplanques2020ecapa} architecture, an enhanced variant of the Time Delay Neural Network (TDNN) with channel-wise correlation matrix attention and time-domain x-vectors. Following the recommended configurations in the WeSpeaker toolkit\cite{wang2023wespeaker}, the SE Res2Block module is set to 512 channels, and the speaker embedding dimension is 192. We do not utilize any TTS-synthesized data during training, nor do we apply score normalization or large-margin fine-tuning.

\subsection{TTS Models}

In our experiments, we evaluated three widely used TTS models, CosyVoice, MaskGCT, and NaturalSpeech 3, to analyze their respective impacts on the recognition of speaker short-utterance task. 

\subsubsection{CosyVoice}
CosyVoice\cite{du2024cosyvoice} is a scalable multilingual zero-shot TTS system that employs supervised semantic tokens from a multilingual ASR model to enhance content consistency and speaker similarity. Its architecture includes a text encoder, a speech tokenizer, a large language model (LLM), and a conditional flow matching module. By integrating x-vectors, it separates content, speaker identity, and prosody—enabling more natural, diverse synthesized speech.

\subsubsection{MaskGCT}
MaskGCT\cite{wang2024maskgct} is a fully non-autoregressive TTS model that removes the need for explicit text–speech alignment or phone-level duration prediction. It follows a two-stage process: first predicting semantic tokens from text, then predicting acoustic tokens based on the semantic tokens. Adopting a mask-and-predict paradigm, MaskGCT generates the tokens in parallel, resulting in high-quality, natural-sounding speech.

\subsubsection{NaturalSpeech 3}
NaturalSpeech 3\cite{ju2024naturalspeech} is an advanced zero-shot TTS system designed to improve speech quality, similarity, and prosody. It decomposes speech into separate subspaces representing content, prosody, timbre, and acoustic details. Using a neural codec with factorized vector quantization (FVQ), it disentangles the speech waveform and a factorized diffusion model generates each attribute independently. This approach enables NaturalSpeech 3 to produce more natural and expressive speech.

\section{Experimental Results}

Table~\ref{tab:eer_comparison} presents the overall experimental results. In this chapter, we will analyze the results from four aspects: TTS models, speech durations, text content, and fusion techniques.

\begin{table}[!htbp]
\centering
\caption{EER Performance Comparison Across TTS Architectures and Fusion Strategies (Evaluated using ECAPA-TDNN on VoxCeleb1)}
\label{tab:eer_comparison}
\adjustbox{max width=0.5\textwidth}{ 
\begin{tabular}{llcccc}
\toprule
\textbf{TTS Model} & \textbf{Fusion Method} & \multicolumn{4}{c}{\textbf{EER (\%) by Speech Duration}} \\ 
\cline{3-6}
 & & 0.5s & 1s & 2s & All \\ \midrule

\textbf{Baseline} & Original Speech & 26.83 & 12.82 & 5.39 & 2.05 \\ \midrule

\multirow{4}{*}{\textbf{NaturalSpeech 3}}
 & TTS Only & 27.13 & 21.38 & 16.58 & 11.72 \\
 & Addition & 22.75 & 12.82 & 7.10 & 3.57 \\
 & Attention & 23.02 & 12.37 & 6.48 & 3.10 \\
 & Weighted Mean & \textbf{22.75} & 11.94 & 5.29 & 2.05 \\\midrule

\multirow{4}{*}{\textbf{CosyVoice}}
 & TTS Only & 27.60 & 15.81 & 8.72 & 3.92 \\
 & Addition & 22.98 & 11.36 & 5.43 & 2.43 \\
 & Attention & 23.08 & 11.16 & 5.19 & 2.26 \\
 & Weighted Mean & 22.98 & 11.05 & 4.96 & 2.02 \\ \midrule

\multirow{4}{*}{\textbf{MaskGCT}}
 & TTS Only & 40.80 & 28.12 & 10.63 & 3.17 \\
 & Addition & 29.39 & 14.46 & 5.08 & 1.90 \\
 & Attention & 28.72 & 13.84 & 4.91 & 1.88 \\
 & Weighted Mean & 25.97 & 11.88 & 4.57 & \textbf{1.83} \\ \midrule

\textbf{CosyVoice+MaskGCT} & Weighted Mean & 23.00 & \textbf{10.77} & \textbf{4.49} & 1.95 \\ \bottomrule
\end{tabular}
} 
\end{table}

\begin{table}[htbp]
\centering
\caption{Phonemic Completeness Analysis in Text Prompts (CosyVoice TTS, 2s Prompts)}
\label{tab:phoneme_impact}
\begin{tabular}{lcc}
\toprule
\textbf{Text Configuration} & \textbf{Phoneme Count} & \textbf{EER (\%)} \\ \midrule
Minimal Phoneme Set & 7 & 13.73 \\
Basic Coverage & 16 & 9.64 \\
Extended Coverage & 26 & 9.23 \\
Full Coverage & 42 & 8.63 \\
Original Transcript & - & 8.66 \\ \bottomrule
\multicolumn{3}{l}{\footnotesize \textit{Note}: Full inventory contains 42 English phonemes per CMUdict} \\
\end{tabular}
\end{table}

\subsection{Comparison of Different TTS Models}

\begin{figure}[htbp]
  \centering
  \includegraphics[width=0.45\textwidth]{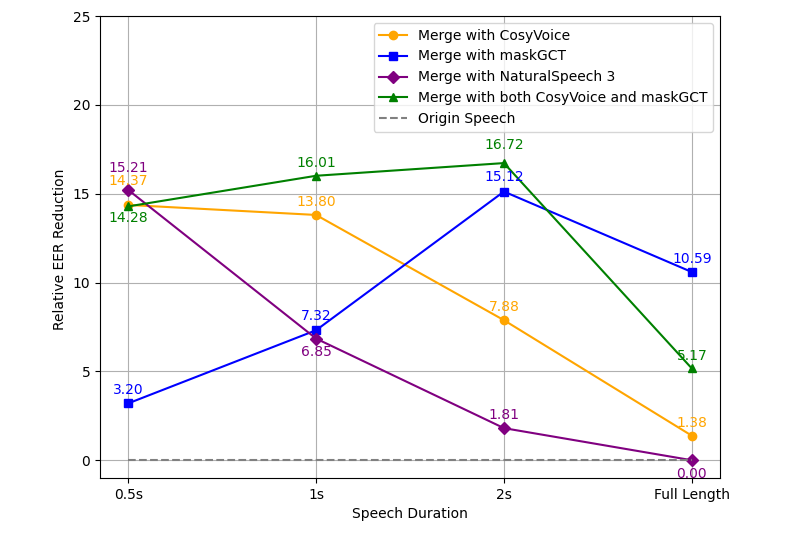}
  \caption{Relative EER Reduction (\%) for Different TTS Models across Various Durations (Fusion with weighted mean)}
  \label{fig:figure2}
\end{figure}

Table~\ref{tab:eer_comparison} and Figure~\ref{fig:figure2} illustrate the impact of TTS-synthesized speech on speaker recognition. We evaluated CosyVoice, MaskGCT, and NaturalSpeech 3 with varying prompt durations. The results show distinct performance patterns: CosyVoice and NaturalSpeech 3 perform better with shorter prompts, while MaskGCT excels with longer ones.


To assess whether these gains stem solely from the increased utterance length introduced by TTS-generated speech, we conducted an experiment where short utterances were repeated to 15 seconds and fused with the original speaker embedding. Results confirm that the primary benefits arise from additional speaker-specific information in TTS-generated speech, rather than duration alone.

\subsection{Impact of the Target Text}


Table~\ref{tab:phoneme_impact} illustrates the impact of text content on speaker recognition performance. In this experiment, we controlled the TTS-synthesized text length while varying its phoneme diversity to assess its influence on verification accuracy. The results indicate that as phoneme coverage decreases, the Equal Error Rate (EER) increases, highlighting the importance of diverse phoneme representation in synthetic speech for improved performance.


Furthermore, we evaluated two text selection strategies: using the original transcript for each utterance versus maintaining a consistent synthesized text across all test speech. Results across three TTS models indicate that text uniformity has minimal impact on performance, suggesting that both original transcripts and consistent test set content are viable choices.

\subsection{Impact of Speech Duration}



Table~\ref{tab:eer_comparison} also presents the equal error rate (EER) for different TTS models across various speech durations. The results confirm that as speech duration decreases, the EER increases significantly for all models, highlighting the considerable challenge of short-utterance speaker recognition.

Notably, our proposed method exhibits a more pronounced relative improvement for very short speech, aligning with our hypothesis that TTS augmentation is particularly beneficial in this scenario. Specifically, for durations less then 2 seconds, incorporating TTS-generated speech leads to an approximate 15\% improvement in performance compared to the baseline. This suggests that TTS-generated speech provides additional speaker-discriminative information when only limited speech is available. However, as speech duration increases, the relative benefit of TTS augmentation diminishes. Additionally, we observe a decline in TTS stability when only short prompts are available, which may impact its effectiveness in certain cases.

\subsection{Comparison of Different Fusion Methods }

In this section, we present a comparative analysis of various fusion methods, with performance differences summarized in Table~\ref{tab:eer_comparison}. Our findings indicate that directly replacing original short utterances with TTS-synthesized speech leads to an increased equal error rate (EER), confirming that TTS-generated speech cannot simply and directly replace real speech in short-duration speaker verification.

However, when TTS-generated speaker embeddings are fused with those from the original speech, system performance improves consistently across all four evaluated fusion strategies.This suggests that TTS-generated speech captures complementary speaker information, enhancing speaker verification in short-utterance scenarios.


\begin{figure}[tbp]
  \centering
  \includegraphics[width=0.47\textwidth]{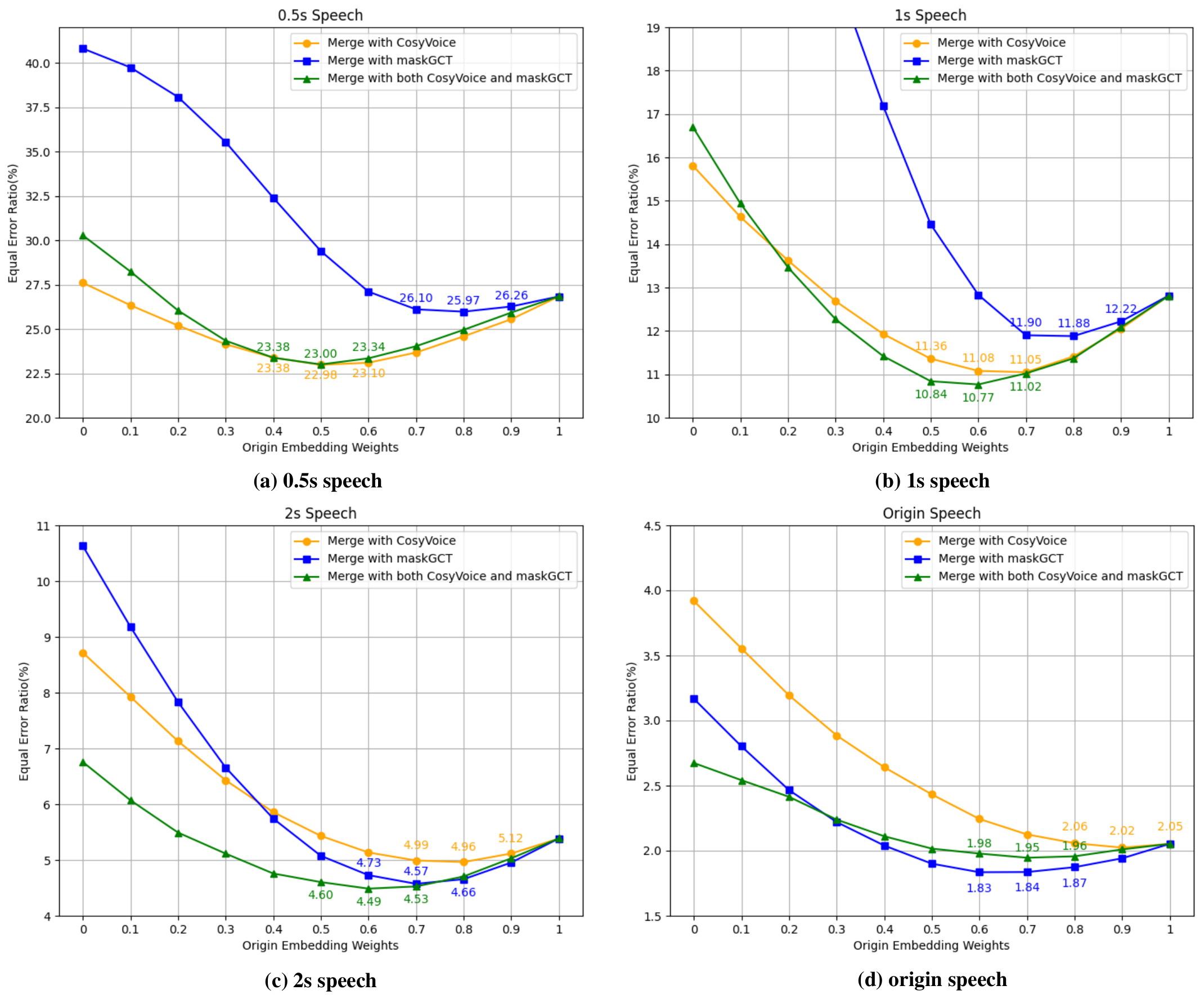}
  \caption{Comparison of EER Performance for Different TTS Models with Varying Origin Embedding Weights across Different Speech Durations (Fusion with weighted mean)}
  \label{fig:figure3}
\end{figure}

Our experimental results demonstrate that, among the fusion methods tested, manual weighted fusion yields the best performance, followed by attention-based fusion. Direct addition ranks third in terms of performance, while simple concatenation shows the least improvement.

Expanding on the insights from Section 4.2, this study further investigates a two-stage fusion process, where embeddings from both MaskGCT and CosyVoice are initially combined, followed by integration with the original speaker embedding. This approach yields the best overall performance, demonstrating reductions in the equal error rate (EER) across all durations compared to the baseline. Additionally, this fused model demonstrates greater consistency in optimal fusion weight selection across different segment lengths. For instance, with an optimal fusion weight of $w=0.6$, the relative EER reductions are 3.6\%, 16.7\%, 16.0\%, and 13.0\% for full-length, 2s, 1s, and 0.5s segments, respectively. This suggests that under this fusion framework, the optimal fusion weight becomes more stable across varying speech durations, enhancing the model’s robustness. Figure~\ref{fig:figure3} provides a more intuitive illustration of the performance improvement during the fusion process.


Finally, we extend our experiments by synthesizing multiple utterances from the same prompt while varying the text content, then fusing three such generated speeches. This approach yields additional performance gains, indicating that even minor variations in synthesized content can introduce complementary speaker cues.

\section{Conclusions}
This paper investigates the potential benefits of zero-shot TTS (ZS-TTS) for short-utterance speaker verification. Through experiments on TTS models, speech durations, text content, and fusion techniques, we observe that while ZS-TTS provides measurable improvements, it also has notable limitations. Directly replacing original speech with TTS-generated speech degrades performance, whereas embedding fusion enhances recognition, particularly for short utterances. We hope these findings offer new insights and contribute to the development of more effective TTS-augmented speaker verification systems.


\bibliographystyle{IEEEtran}
\bibliography{mybib}

\end{document}